\documentclass[twocolumn,preprintnumbers,amssymb,amsmath,aps,prx]{revtex4-1}

\usepackage{tabularx}
\usepackage{float}
\usepackage{bm}
\usepackage{graphicx}
\usepackage{amsmath}
\usepackage{color}

\begin{document}
	\title{Topological Metamaterials based on polariton rings}
	
	\author{V. K.~Kozin$^{1,2}$}\email{vak13@hi.is}
	\author{I. V.~Iorsh$^{2}$}\email{i.iorsh@phoi.ifmo.ru}
	\author{A. V.~Nalitov$^{1,2}$}
	\author{I. A.~Shelykh$^{1,2}$}
	
	\affiliation {$^1$Science Institute,
		University of Iceland, Dunhagi-3, IS-107 Reykjavik, Iceland}
	\affiliation {$^2$ITMO University, St. Petersburg 197101, Russia}
	
	\pacs{to be filled.}
	
	\begin{abstract}
		Chern insulator phase is shown to emerge in two dimensional arrays of polariton rings where time-reversal symmetry is broken due to the application of out-of-plane magnetic field. The interplay of Zeeman splitting with the photonic analog of spin-orbit coupling (TE-TM splitting) inherently present in this system leads to the appearance of synthetic U(1) gauge field and opening of topologically nontrivial spectral gaps. This results in the onset of topologically protected chiral edge states similar to those forming in quantum Hall effect. In one dimensional zigzag arrays of polariton rings edge states similar to those appearing in Su-Schrieffer–Heeger (SSH) model are formed.
	\end{abstract}
	\maketitle
	
	\section{Introduction}
	
	Since the discovery of the quantum Hall effect~\cite{KlausvonKlitzing} and its interpretation in terms of topological invariance~\cite{PhysRevLett.49.405}, there were several breakthroughs in the  field of topological condensed matter physics. Fractional quantum Hall effect~\cite{PhysRevLett.48.1559,Laughlin_argument} was interpreted in terms of composite fermion theory by Jain~\cite{JainBook} and non-Abelian anyon statistics by Xiao-Gang Wen~\cite{WenAnyon}.
	Later on, ideas of the topological protection of the quantum Hall phase edge states were generalized to the concept of the bulk-boundary correspondence ~\cite{XIAOGANGWEN1994476}.
	Finally, introduction of the hierarchy of topological invariants~\cite{Z_2_top_order} drastically increased the range of available topologically nontrivial electronic configurations and corresponding edge states~\cite{ZHasan}.
	
	Recent decade has seen the rise of topological photonics, following the prediction of topologically protected optical crystal edge states similar to the conducting electronic edge states~\cite{phot_cryst_broken_trs}.
	Existing proposals for topological photonics exploit symmetry breaking with synthetic magnetic fields in arrays of coupled waveguides~\cite{Hafezi2011}, spatial analogue of Floquet modulation~\cite{Rechtsman2013}, and magneto-optic metamaterials~\cite{Khanikaev2012}.
	Recently, strong light-matter interaction in coupled microcavities was predicted to yield topological polaritonic edge states~\cite{NalitovZ_TI_2015,Topolaritons,PhysRevB.91.161413}.
	
	In contrast to topological photonics, the topological states of polaritonic systems can be controlled with real magnetic field or via strong particle- particle interactions~\cite{PhysRevB.93.085438}. The latter give additional twist to polariton based systems, which can demonstrate nonlinear topological effects related to optical bistability~\cite{PhysRevLett.119.253904}.     Moreover, bosonic stimulation of polariton-polariton scattering allows spontaneous coherent emission from topologically nontrivial states~\cite{St-Jean2017}.
	Overall, the nonlinearity stemming from polariton-polariton interactions, responsible for the crossover from polariton lasing to polariton Bose-Einstein condensation, provides a unique opportunity of studying new interacting bosonic topological phases.
	
	Polaritons are neutral particles, so the application of an external magnetic field affects only their spin but not orbital motion. However, as it was shown in the Ref.\cite{ShelykhSoliton} in non-simply connected geometry such as polariton ring, the  interplay of Zeeman splitting with the photonic analog of spin-orbit coupling (TE-TM splitting) inherently present in this system leads to the appearance of synthetic U(1) gauge field affecting orbital motion. The effect is due to the appearance of non-zero geometric Berry phase during one round of the rotation along the ring \cite{ShelykhBerry}. In this paper we extend the idea of synthetic U(1) field for polariton system to the case of periodic arrays of microcavity rings. We demonstrate that in a 2D array of polariton rings the presence of synthetic U(1) field leads to the nontrivial band topology, characterized by nonzero Chern numbers, and induces topologically protected unidirectional edge states, similar to those appearing in QHE. In 1D zigzag array of the rings edge states similar to those appearing in Su-Schrieffer-Heeger (SSH) model are formed.
	
	The work is organized as follows. In Section II we construct and diagonalize the Hamiltonian of a single polariton ring accounting for both TE-TM and Zeeman splitting. The results are then used for the topological analysis of two-dimensional cavity ring arrays presented in Section III and zigzag chains of annular cavities presented in Section IV. Conclusions summarize the results of the work.
	
	\section{Single polariton ring}
\begin{figure}[h!]
		\centering
		\includegraphics[width=0.45 \textwidth]{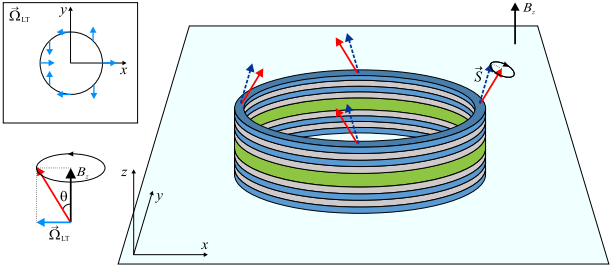}
		\caption{Schematic of the considered geometry. The cavity (polariton ring) is constituted by two Distributed Bragg Reflectors sandwiching a cavity with embedded
    Quantum Well. The polariton ring
is placed into external magnetic-field $B_z$ perpendicular to its interface.
The total effective magnetic field acting on the polariton’s spin is
a combination of the real magnetic field and the field provided by
TE-TM splitting. The direction of the total effective magnetic field
changes along the ring as is shown by the red solid arrows. If one moves along
the ring it twice covers a cone characterized by angle $\theta$. The dashed arrow shows the direction of the Stokes vector $\vec{S}$. In adiabatic approximation the direction of the Stokes vector coincides with the direction of the total effective magnetic field, while they become different for the exact solution of the Schrodinger equation.}
		\label{fig:Single_ring}
	\end{figure}

	In this section we derive formally the effective 1D Hamiltonian describing a single polariton ring (see Fig.~\ref{fig:Single_ring}) with the TE-TM splitting in the presence of a magnetic field. 
	We start with from the Hamiltonian of 2D polariton inside a planar microcavity \cite{FlayacVortex} 
	\begin{equation}\label{eq:H_2D}
	\widehat{H}_{2D}= 
	\begin{pmatrix}
	\widehat{H}_0(\hat{\mathbf{k}})+\Delta_z/2 & \widehat{H}_{\text{TE-TM}}(\hat{\mathbf{k}}) \\
	\widehat{H}^{\dagger}_{\text{TE-TM}}(\hat{\mathbf{k}}) & \widehat{H}_0(\hat{\mathbf{k}})-\Delta_z/2
	\end{pmatrix},
	\end{equation}
	where the diagonal terms $\hat{H}_0$ describe the kinetic energy
	of lower cavity polaritons, and the off-diagonal terms
	$\widehat{H}_{\text{TE-TM}}$ correspond to the longitudinal-transverse splitting. We further employ the effective
	mass approximation  
	\begin{equation}
	\widehat{H}_0(\hat{\mathbf{k}})=\frac{\hbar^2 \hat{\mathbf{k}}^2}{2 m_{\text{eff}}}.
	\end{equation}
	The TE-TM part is given by
	\begin{equation}
	\widehat{H}_{\text{TE-TM}}(\hat{\mathbf{k}})=\beta\left(\frac{\partial}{\partial y}+i\frac{\partial}{\partial x}\right)^2,
	\end{equation}
	where $\beta$ is a constant, characterizing the strength of the TE-TM
	splitting which can be expressed via the longitudinal and
	transverse polariton effective masses $m_l$ and $m_t$ as $\beta=(\hbar^2/4)(m_l^{-1}-m_t^{-1})$. In order to proceed with the derivation of the correct 1D Hamiltonian let us pass to the cylindrical coordinates and add the confining potential $V(r)$, which forces the polariton wave functions to be localized on the ring in the radial direction to $\widehat{H}_{2D}$. The terms associated with the TE-TM splitting rewritten in polar coordinates read	
	\begin{equation}\label{TETM_cyl}
	\begin{aligned}
	&\left(\frac{\partial}{\partial y}\pm i\frac{\partial}{\partial x}\right)^2=e^{\mp 2 i\varphi}\times\\
	&\left(-\frac{\partial^2}{\partial r^2}\pm \frac{2 i}{r}\frac{\partial^2}{\partial r \partial \varphi}\mp  \frac{2 i}{r^2}\frac{\partial}{\partial\varphi}+\frac{1}{r}\frac{\partial}{\partial r}+\frac{1}{r^2}\frac{\partial^2}{\partial \varphi^2}\right).
	\end{aligned}
	\end{equation}
	We decompose the Hamiltonian Eq.~(\ref{eq:H_2D}) into two parts $\widehat{H}_{2D}=\widehat{H}_0(r)+\widehat{H}_1(r,\varphi)$ where
	\begin{equation}\label{H_0}
	\widehat{H}_0(r)=-\frac{\hbar^2}{2m_{\text{eff}}}\left(\frac{\partial^2}{\partial r^2}+\frac{1}{r}\frac{\partial}{\partial r}\right)+V(r).
	\end{equation}
	Following the conventional procedure~\cite{correct1DHam}, the Hamiltonian of  1D ring is given by
	\begin{equation}
	\widehat{H}=\langle R_0(r)\lvert\widehat{H}_1(r,\varphi)\rvert R_0(r)\rangle,	
	\end{equation}
	where $R_0(r)$ is a the lowest radial mode of the Hamiltonian (\ref{H_0}). After averaging all the terms in Eq.~(\ref{TETM_cyl}) (see details in Appendix~\ref{appendix:der_corr_ham}) we arrive at 
	\begin{equation}\label{ring_ham}
	\widehat{H}= 
	\frac{\hbar^2}{2m_{\text{eff}}R^2}\begin{pmatrix}   \hat{\tilde{k}}^2+\Delta_2/2 & \Delta_1e^{-2i\varphi} \\
	\Delta_1e^{2i\varphi} & \hat{\tilde{k}}^2-\Delta_2/2
	\end{pmatrix},
	\end{equation}
	where $\hat{\tilde{k}}=-i (d/d\varphi)$, $R$ is the radius of the ring.  For the sake of simplicity we introduced dimensionless parameters $\Delta_{1,2}$ corresponding to the LT and Zeeman splittings as $\Delta_{\mathrm{LT}(z)}=\Delta_{1(2)}\hbar^2/(2m_{\text{eff}} R^2)$ and $\Delta_{LT}\approx\beta/2a^2$ with $a$ being lateral width of the ring (see Appendix A for the derivation). The Hamiltonian~(\ref{ring_ham}) was proposed earlier basing on symmetry considerations~\cite{ShelykhBerry}. The approach developed here allows to establish correspondence between the parameters of the Hamiltonian and the geometrical dimensions of the structure.
	
	In general, solutions of the stationary Schrodinger equation with Hamiltonian~(\ref{ring_ham}) can be represented in the following form:
	\begin{equation}\label{eq:ring_eigenfunc}
	\widetilde{\psi}(\varphi)=\widetilde{\chi}(\varphi,\tilde{k}) e^{i \tilde{k}\varphi},
	\end{equation}
	where $\widetilde{\chi}(\varphi,\tilde{k})$ is a corresponding spinor
	\begin{equation}
	\widetilde{\chi}(\varphi,\tilde{k}) =\frac{1}{\sqrt{\xi(\tilde{k})^2+1}}\left(
	\begin{array}{c}
	e^{-i\varphi}\\
	\xi(\tilde{k})e^{i\varphi}\\
	\end{array}
	\right),
	\end{equation}
	and $\xi(\tilde{k})=\Delta_1/((E+\Delta_2/2)-(1+\tilde{k})^2)$. The energy of the state Eq.~(\ref{eq:ring_eigenfunc}) (measured in units of $\hbar^2/2m_{\text{eff}}R^2$) is given by:
	\begin{equation}\label{ring_disp}
	E=(\tilde{k}^2+1)\pm\sqrt{\Delta_1^2+(\Delta_2/2-2\tilde{k})^2},
	\end{equation}
	For a given energy Eq.~(\ref{ring_disp}) has four solutions for $\tilde{k}_\text{i}$, $\text{i}=1\ldots4$ two positive and two negative, corresponding to clockwise and anti-clockwise propagation and two opposite spin orientations. Analytical expressions for them are listed in the Appendix~\ref{appendix:disp_sing_ring}. It should be noted that if the Zeeman splitting is present, $E(\tilde{k})\neq E(-\tilde{k})$ which corresponds to the breaking of the time-reversal symmetry in the system leading to non-equivalence of clockwise and anti-clockwise propagation directions. In the arrays of interconnected rings this will lead to the appearance of the topologically protected edge states as we will demonstrate later on.
	
	The general solution corresponding to a given energy thus reads: 
	\begin{equation}\label{eq:ring_gen_sol}
	\widetilde{\Psi}(\varphi)=\sum\limits_{\text{i}}\widetilde{C}^{(\text{i})}\widetilde{\psi}_\text{i}(\varphi),
	\end{equation}
	where the summation is performed over all real roots of Eq.~(\ref{ring_disp}) and $\widetilde{\psi}_\text{i}(\varphi)$ is an eigenfunction (Eq.~(\ref{eq:ring_eigenfunc})) at $\tilde{k}=\tilde{k}_\text{i}$. The Stokes vector given by
	\begin{equation}
	    \vec{S}=\widetilde{\psi}^{\dagger}\vec{\sigma}\widetilde{\psi}=
	    \left(
    	\begin{array}{c}
        	\sin{\theta}\cos{2\varphi}\\
        	\sin{\theta}\sin{2\varphi}\\
        	\cos{\theta}
    	\end{array}
    	\right).
	\end{equation}
	characterizes the distribution of the spin projection along the ring and defines the profile of the polarization of the emission along the ring, and ${\tan{\theta/2}=\xi(\tilde{k})}$. Note, that the direction of the Stokes vector is not exactly the same as direction of the total effective magnetic field as in the case of adiabatic approximation \cite{ShelykhBerry}.
	
	As we consider only bright excitons with spin $\pm1$ as a two-level system, the z-projection of the operator of total angular momentum is $\hat{J}_z=\hbar\hat{\tilde{k}}+\hbar\sigma_z$. One can check that $\hat{J}_z\widetilde{\psi}(\varphi)=\hbar \tilde{k}\widetilde{\psi}(\varphi)$ which clarifies the physical meaning of $\tilde{k}$. If there is only a single isolated ring, one should impose periodic boundary condition $\widetilde{\psi}(\varphi)=\widetilde{\psi}(\varphi+2\pi)$, so that $\tilde{k}$ is integer and corresponds to the orbital quantized angular momentum and according to Eq.~(\ref{ring_disp}) energy becomes quantized as well. 
	
	\section{Two-dimensional array of rings}
\begin{figure}[h]
		\includegraphics[width=0.35\textwidth]{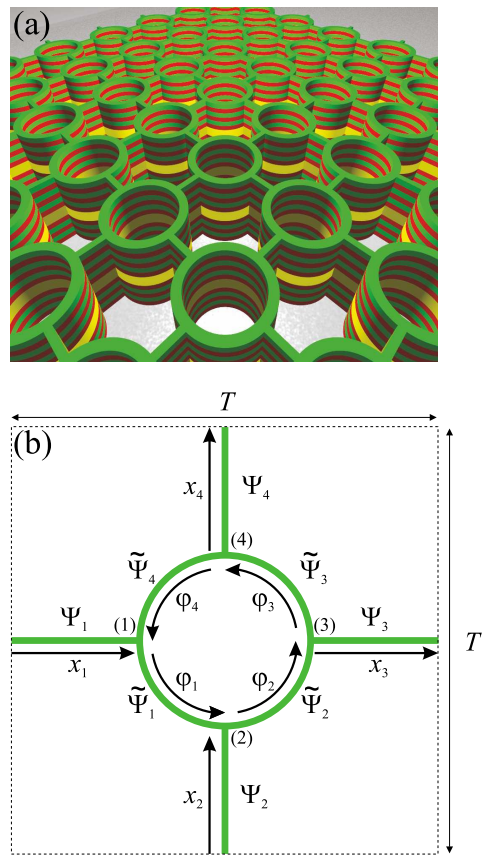}
		\caption{(a) Geometry of the structure: two-dimensional square lattice of the exciton-polariton ring resonators connected via leads. (b) A unit cell of the structure: a polariton ring of radius $R$ with four attached leads of length $d/2$. The period of the structure is $T=d+2R$, and the corresponding variables are defined as $x_{1,2}\in[0,d/2]$, $x_{3,4}\in[-d/2,0]$ and $\varphi_{1,\ldots,4}\in[0,\pi/2]$. The ring-lead junctions are labeled as (1)-(4).}
		\label{fig:sketch_pl}
	\end{figure}

	We consider a two-dimensional array of polariton rings, as shown in Figs.~\ref{fig:sketch_pl} (a) and (b). Each ring can be considered as a plaquette in a square lattice where the wave propagates via leads connecting neighboring cavity rings. Every ring has four leads attached to it. There is an applied magnetic field $B_z$ in addition to the effective magnetic field $\vec \Omega_{\mathrm{LT}}$ that stems from the TE-TM splitting and lies in the plane of the array. The splitting values $\Delta_z = g\mu_\mathrm{B} B_z$, $\Delta_{\mathrm{LT}} = g\mu_\mathrm{B} |\vec{\Omega}_{\mathrm{LT}}|$, where $g$ is the effective Lande $g$-factor for the 2D exciton and $\mu_\mathrm{B}$ is the Bohr magneton. In the unit cell presented in Fig.~\ref{fig:sketch_pl}(b) the wave function is piecewise-defined.
	
	\begin{figure*}[t]
		\centering
		\includegraphics[width=1\textwidth]{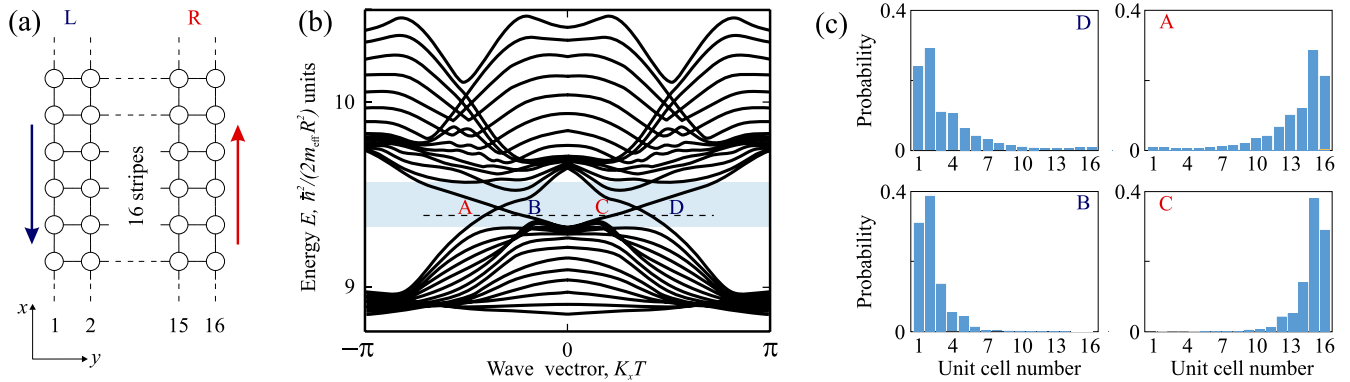}
		\caption{Topologically protected edge states for $\Delta_1=5$, $\Delta_2=2$ and $h=d/R=1.2$. The corresponding Chern number $C=2$ is a sum of all $C_n$ below the energy level (denoted by the dashed black line). (a) The geometry of the structure: a two-dimensional square lattice of the exciton-polariton ring resonators that is infinite in the $x$-direction and finite in the $y$-direction. Blue and red arrows indicate the directions of the propagation along left and right edges. (b) Band structure of the stripe consisting of 16 periods in the $y$-direction. The rectangle marks the bandgap of the bulk system. For a given energy marked by the dashed line there are two edge states labeled as A and C corresponding to the right boundary, and two states B and D corresponding to the left boundary. Note, that the sign of the group velocity is the same for the states propagating along the same edge. This means that these states are chiral and topologically protected with respect to backscattering. (c) Probability distribution profiles of the edge states marked at the band diagram. Direction of their propagation is set by the sign of the product $B_z g$. }
		\label{fig:top_states}
	\end{figure*}
	
	For the lead (polariton wire), the wave function can be obtained similarly to the case for a ring  by noticing that the polariton wire Hamiltonian is the ring Hamiltonian (Eq.~(\ref{ring_ham})) at the limit $R\to\infty$.  For example, two branches (upper and lower, due to the TE-TM and Zeeman splittings) of the dispersion are given by 
	\begin{equation}\label{eq:lead_disp}
	E=(k R)^2\pm\sqrt{\Delta_1^2+(\Delta_2/2)^2},
	\end{equation} 
	where again $E$ is measured in $\hbar^2/(2 m_{\text{eff}} R^2)$ units.
	When $E\ge\sqrt{\Delta_1^2+(\Delta_2/2)^2}$ we have four eigenfunctions of a lead
	
	\begin{equation}
	\psi_{1,2}(x)=\frac{1}{\sqrt{\xi_0^2+1}}\left(
	\begin{array}{c}
	-\xi_0\\
	1\\
	\end{array}
	\right)e^{\pm i k_l x},
	\end{equation}
	
	\begin{equation}
	\psi_{3,4}(x)=\frac{1}{\sqrt{\xi_0^2+1}}\left(
	\begin{array}{c}
	1\\
	\xi_0\\
	\end{array}
	\right)e^{\pm i k_u x},
	\end{equation}
	where $k_{u,l}=R^{-1}\sqrt{E\mp\sqrt{\Delta_1^2+(\Delta_2/2)^2}}$ and
	\begin{equation}
	\xi_0=\frac{\Delta_1}{\Delta_2/2+\sqrt{\Delta_1^2+(\Delta_2/2)^2}}.
	\end{equation}
	In the case $E<\sqrt{\Delta_1^2+(\Delta_2/2)^2}$ only $\psi_{1,2}(x)$ are relevant. Hence, the general solution for the $\text{m}$-th lead ($\text{m}=1\ldots4$, see Fig.~\ref{fig:sketch_pl}(a)) read
	\begin{equation}
	\Psi_\text{m}(x)=\sum\limits_{\text{i}}C_\text{\text{m}}^{(\text{i})}\psi_\text{i}(x),
	\end{equation}
	where the summation is performed over all real roots of Eq.~(\ref{eq:lead_disp}) for a given energy.
	Similarly, for the $\widetilde{\text{m}}$-th arc ($\widetilde{\text{m}}=1\ldots4$), in obedience to Eq.~(\ref{eq:ring_gen_sol}), we have
	\begin{equation}
	\widetilde{\Psi}_{\widetilde{\text{m}}}(\varphi)=\sum\limits_{\text{i}}\widetilde{C}_{\widetilde{\text{m}}}^{(\text{i})}\widetilde{\psi}_\text{i}(\varphi),
	\end{equation}
	where the summation is performed over all real roots of Eq.~(\ref{ring_disp}).
	
	To define the behaviour of the polariton waves at the junctions connecting the rings and the leads we use Griffith's boundary conditions~\cite{Griff} stating that the wave functions have to be continuous and the input probability currents must be exactly compensated by the output. Together with the Bloch periodic boundary condition this gives the closed set of linear algebraic equations allowing to define the band structure of the system (see Appendix~\ref{appendix:2D_array_sys_derivation} for the details). It should be noted that one may use the $S$-matrix approach instead of imposing Griffith's conditions (see Appendix~\ref{appendix:s-matrix}, where we present the further development of the $S$-matrix theory).
	
	Once the spectrum of the bulk system is obtained from the secular equation, the Chern numbers can be calculated. 
	The Chern number corresponding to the $n$-th band of the considered periodic array is defined as
	\begin{equation}\label{ChN}
	C_n=\frac{1}{2\pi i}\int\limits_{1\text{BZ}}d^2 K F_{xy}(\mathbf{K}),
	\end{equation}
	where $F_{xy}(\mathbf{K})={\partial A_y}/{\partial K_x}-{\partial A_x}/{\partial K_y}$ is the field strength associated with the Berry connection, $A_j(\mathbf{K})=\langle n(\mathbf{K})|\frac{\partial}{\partial K_j}|n(\mathbf{K})\rangle$
	is the vector potential of the field, and $|n(\mathbf{K})\rangle$
	is the normalized Bloch wave function of the $n$-th
	band~\cite{Bernevig_book}. The integration is performed over the first Brillouin zone (1BZ). 
	
	Topological gaps open in the case of non-zero  both TE-TM and Zeeman splittings. The Chern numbers are calculated for the gapped system (see the caption of Fig.~\ref{fig:top_states}). Since there are gaps with non-zero Chern invariant, the bulk-boundary correspondence suggests the existence of topologically protected edge states for a finite lattice. To explore the properties of the edge states, one should add the boundaries to the system. Let us consider the system depicted in Fig.~\ref{fig:top_states}(a): a strip composed of 16 rings in the $y$-direction and infinite in the $x$-direction. The spectrum of the strip can be obtained by writing down Eqs.~(\ref{eq:QPC1})-(\ref{eq:QPC4}) for the rings $2\ldots15$ supplemented by a slightly modified version of these equations for the outer rings (number 1 and 16 in Fig.~\ref{fig:top_states}(a)) as they are deprived of the first and fourth leads respectively. Also, one should add the Bloch condition only for the $x$-direction as the periodicity is broken in the $y$-direction. Finally, we obtain a linear system for the variables $\{C_{\aleph,\text{m}}^{(\text{i})}\}\cup\{ \widetilde{C}_{\aleph,\widetilde{\text{m}}}^{(\text{i})}\}$ (where $\aleph=1\ldots16$ numbers the unit cells along an arbitrarily chosen row in Fig.~\ref{fig:top_states}(a)), with the secular equation yielding the spectrum, presented in Fig~\ref{fig:top_states}(b). 
	
	As can be seen from Fig.~\ref{fig:top_states}(b), there exist topological edge states at the bulk gap. For a given energy level, plotted with the dashed black line in Fig.~\ref{fig:top_states}(b), there exist two edge states per boundary. The wavefunction density at the unit cell scale, corresponding to the edge states, is shown in Fig.~\ref{fig:top_states}(c). Similar to the case of QHE, edge states are chiral and topologically protected: direction of the propagation is linked to the edge, so backscattering is possible only if it is accompanied with hopping from one edge to another.  

	\section{Zigzag chain of polariton rings}
\begin{figure}[ht]
		\centering
		\includegraphics[width=0.35 \textwidth]{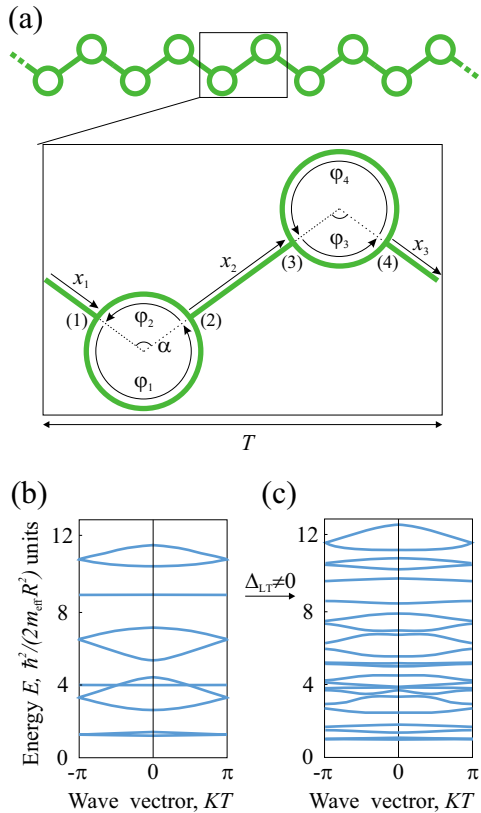}
		\caption{(a) Sketch of a zigzag array of polariton rings and a unit cell of the structure. Each ring is of radius $R$ and each lead is of length $d$. The period of the structure is $T=2(d+2R)\sin{\alpha/2}$. The corresponding variables are defined as $\varphi_{1,4} \in [0,2\pi-\alpha]$, $\varphi_{2,3} \in [0,\alpha]$ and $x_1 \in [0, d/2]$, $x_2 \in [-d/2, d/2]$, $x_3 \in [-d/2, 0]$. The band diagram for an infinite zigzag chain (with $\alpha=\pi/2$) in the absence of LT-splitting ($\Delta_1=0$) (b) and at $\Delta_1=2$ (c).}
		\label{fig:SSH}
	\end{figure}

	We now proceed to the analysis of zigzag arrays of polariton rings. Let us consider a system of rings connected into a zigzag chain as shown in the  Fig.~\ref{fig:SSH}. It is reminiscent of the Su-Schrieffer-Heeger model~\cite{ZHasan}, where each carbon atom of polyacetylene is replaced with a polariton ring and the role of the `relative-bond-strength' (for polyacetylene) is played by the `bond-angle' $\alpha$  (see Fig.~\ref{fig:SSH}(a)). In this section we focus on the effects that arise in the absence of an external magnetic field ($B_z=0\Rightarrow\Delta_2=0$). The application of the external magnetic field in the case of 1D chain of the rings does not change significantly the results.
	
	As shown in Fig.~\ref{fig:SSH}(a), there are two rings per unit cell and within each unit cell there are different effective magnetic fields in the leads that are not parallel to each other. The directions of these effective magnetic fields depend on the angle $\alpha$.  
	
	In a similar manner as in the previous section we immediately derive the scattering equations given by Griffith's and Bloch boundary conditions (see the derivation in Appendix~\ref{appendix:zigzag_sys_derivation}).
	
	\begin{figure}[ht]
		\includegraphics[width=0.5 \textwidth]{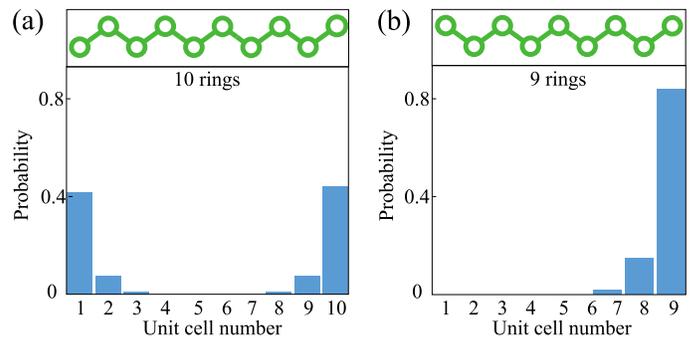}
		\caption{Probability distribution of the edge states in polariton ring zigzag chains with $\alpha=\pi/2$ and $h=d/R=1.2$ for even (a) and odd (b) number of rings. In the odd number case (b) there also exists a symmetric edge state that is localized on the left side of the chain. The profiles are shown for $\Delta_1=2$, $E=8.985$ (in $\hbar^2/(2m_{\text{eff}}R^2)$ units). The energy value is chosen to lay within the bulk band gap.}
		\label{fig:SSH_density}
	\end{figure}
	
	For $\alpha=\pi/2$ and zero effective magnetic field, the corresponding band diagram (see Fig.~\ref{fig:SSH}(b)), exhibits flat bands are reminiscent of Landau levels related to the states localized within the ring. As we introduce a non-zero LT-splitting, new gaps open (see Fig.~\ref{fig:SSH}(c)). As one introduces boundaries to the system the edge states appear (see Fig.~\ref{fig:SSH_density}(a) and (b)). In both cases in Fig.~\ref{fig:SSH_density} (the even and the odd number of the rings), the edge states are twice degenerate. Thus, by taking their linear combination one can prepare a state localized either at both sides of the chain or at one side.  We note that in the case of the chain, no external magnetic field is required for the emergence of the edge-states, since the angle $\alpha$ not equal to $\pi$ breaks the equivalence of the clockwise and counter-clockwise propagating modes in the separate rings. As the angle $\alpha$ gets closer to $\pi$, the localization length of the edge states increases. 
	
	\section{Conclusions and Outlook}
	To conclude, we have shown that a two-dimensional array of the polariton rings is characterized by the non-trivial Chern invariants and supports topologically protected chiral edge states analogously to the case of Quantum Hall insulator. Also, the edge states are shown to emerge in a zigzag chain of polariton rings. Existence of topologically protected states can find its applications in spinoptronics. 
	
	\section*{Acknowledgements}
	The work was partially supported by the RISE Program
    (project CoExAN), the Russian Foundation for Basic Research (Project No. 17-02-00053), Rannis project 163082-051,
    Ministry of Education and Science of Russian Federation
    (Projects No. 3.4573.2017/6.7, No. 3.2614.2017/4.6,
    No. 3.1365.2017/4.6, No. 3.8884.2017/8.9, and No.
    14.Y26.31.0015), and Government of Russian Federation
    (Grant No. 08-08).
	
	\appendix
	
	\section{Derivation of the Hamiltonian~(\ref{ring_ham})}
	\label{appendix:der_corr_ham}
	For the confining potential $V(r)=A(r-R)^2/2$ from Ref.~\cite{correct1DHam} we have
	\begin{equation}
	R_0(r)=\left(\frac{\gamma}{R\sqrt{\pi}}\right)^{1/2}e^{-\gamma^2(r-R)^2/2},
	\end{equation}
	where $\gamma^4=m_{\text{eff}} A/\hbar^2$ and the 1D limit is putting $\gamma$ to infinity. The desirable Hamiltonian $\widehat{H}=\langle R_0(r)\lvert\widehat{H}_1(r,\varphi)\rvert R_0(r)\rangle$ contains three types of terms. Namely, the first type is $\langle R_0(r)\lvert r^{-2}\rvert R_0(r)\rangle$, arising from the kinetic and TE-TM parts of $\widehat{H}_1(r,\varphi)$. The other two terms $\langle R_0(r)\lvert \partial^2/\partial r^2\rvert R_0(r)\rangle$ and $\langle R_0(r)\lvert r^{-1}\partial/\partial r\rvert R_0(r)\rangle$ in turn stem only from the TE-TM part. Due to the fact that $R_0^2(r)=\gamma/(R\sqrt{\pi})e^{-\gamma^2(r-R)^2}$ converges to $\delta(r-R)/R$ as ${\gamma\to\infty}$, the first term yields $\lim\limits_{\gamma\to\infty}\langle R_0(r)\lvert r^{-2}\rvert R_0(r)\rangle=R^{-2}$. The second term can be calculated straightforwardly $\langle R_0(r)\lvert r^{-1}\partial/\partial r\rvert R_0(r)\rangle=1/2R_0^2(r)\vert_0^{\infty}=0$.  
	
	The direct calculation of the third term yields 
	$\langle R_0(r)\lvert \partial^2/\partial r^2\rvert R_0(r)\rangle\sim-\gamma^2/2$ when ${\gamma\to\infty}$, so this term dominates over the others in $\langle R_0(r)\lvert\widehat{H}_{\text{TE-TM}}(r,\varphi)\rvert R_0(r)\rangle$ and we can neglect them. Thus, we arrive at the correct Hamiltonian~(\ref{ring_ham}) of a polariton ring with $\Delta_{\mathrm{LT}}=\gamma^2\beta/2$ which, in turn, is inversely proportional to the characteristic width of the microcavity.
	
	\section{Dispersion for an isolated ring}
	\label{appendix:disp_sing_ring}
	In this section we provide the exact analytical solutions of Eq.~(\ref{ring_disp}) for a given energy $E$. For this purpose we introduce the auxiliary variables $a_{1,2}$ and $b$ given by
	\begin{widetext}
		\begin{equation}
		\begin{aligned}
		&a_1=\Bigl(-72 \Delta_1^2+36 \Delta_2^2+64 E^3-96 E^2-72 \Delta_1^2 E-18 \Delta_2^2 E-96 E+64+\\
		&\sqrt{4 \left(4
			(E+1) \left(-9 \Delta_1^2+4 E (2 E-5)+8\right)-9 \Delta_2^2
			(E-2)\right)^2-\left(4 \left(-3 \Delta_1^2+4 (E-1) E+4\right)-3 \Delta_2^2\right)^3}\Bigr)^{1/3},
		\end{aligned}
		\end{equation}
	\end{widetext}
	
	\begin{align}
	&a_2=16-16E+16E^2-12 \Delta_1^2-3 \Delta_2^2,\\
	&b=(4(E+1)+a_2/a_1+a_1)/12.
	\end{align}
	
	As a result, the equations on the dispersion relations for a single polariton ring (Eq.~(\ref{ring_disp})) can be conveniently expressed in these variables:
	
	\begin{align}
	\tilde{k}_{1,2}&=\sqrt{b}\mp\sqrt{E+1-b- \Delta_2/(2\sqrt{b})},\\
	\tilde{k}_{3,4}&=-\sqrt{b}\mp\sqrt{E+1-b+ \Delta_2/(2\sqrt{b})}.
	\end{align}
	
	\section{Derivation of spin current}
	\label{appendix:spin_current}
	The Schrodinger equation for a polariton in a ring, as it follows from the Hamiltonian~(\ref{ring_ham}), reads
	\begin{equation}\label{schrod}
	i\frac{2m_{\text{eff}}R^2}{\hbar}\frac{\partial \Psi}{\partial t}=-\frac{\partial^2\Psi}{\partial\varphi^2}+\Delta_1\sigma_r(2\varphi)\Psi+\frac{\Delta_2}{2}\sigma_z\Psi,
	\end{equation}
	where $\sigma_r(\varphi)=\sigma_x \cos{\varphi}+\sigma_y\sin{\varphi}$ is the Pauli matrix in the cylindrical
	coordinates. The probability current in the ring can be derived from the continuity equation
	\begin{equation}
	\frac{\partial\rho}{\partial t}+\frac{1}{R}\frac{\partial J}{\partial \varphi}=0,
	\end{equation}
	where the probability density is $\rho=\overline{\Psi}^{\intercal}\Psi$ and $\overline{\Psi}$ denotes the complex conjugate of $\Psi$. From the Schrodinger equation~(\ref{schrod}) we have
	\begin{equation}
	\frac{\partial\rho}{\partial t}=\frac{1}{i\hbar}\left(\overline{\Psi}^{\intercal}\widehat{H}\Psi-(\overline{\widehat{H}\Psi})^{\intercal}\Psi\right),
	\end{equation}
	where 
	\begin{equation}
	\widehat{H}=\frac{\hbar^2}{2m_{\text{eff}}R^2}\left(-\frac{\partial^2}{\partial\phi^2}+\Delta_1\sigma_r(2\phi)+\frac{\Delta_2}{2}\sigma_z\right).
	\end{equation}
	Using the following equalities \begin{equation}
	(\overline{\sigma_r(2\phi)\Psi})^{\intercal}\Psi=\overline{\Psi}^{\intercal}\sigma_r(2\phi)\Psi
	\end{equation}
	and
	\begin{equation}
	(\overline{\sigma_z\Psi})^{\intercal}\Psi=\overline{\Psi}^{\intercal}\sigma_z\Psi,
	\end{equation}
	we arrive at
	\begin{equation}
	J=\frac{\hbar}{2m_{\text{eff}}R}\text{Re}\left\lbrace-i\overline{\Psi}^{\intercal}\frac{\partial\Psi}{\partial\varphi}\right\rbrace.
	\end{equation}
	\section{2D array}
	\label{appendix:2D_array_sys_derivation}
	The scattering equations for the junctions (1) through (4) in Fig.~\ref{fig:sketch_pl}(b) are given by Griffith's conditions~\cite{Griff}  and read as follows 
	\begin{align}
	\begin{split}
	&\Psi_1(d/2)=\widetilde{\Psi}_4(\pi/2)=\widetilde{\Psi}_1(0)\\
	&R\Psi^{\prime}_1(d/2)+\widetilde{\Psi}^{\prime}_4(\pi/2)=\widetilde{\Psi}^{\prime}_1(0),
	\end{split}\label{eq:QPC1}\\[10pt]
	\begin{split}
	&\Psi_2(d/2)=\widetilde{\Psi}_1(\pi/2)=\widetilde{\Psi}_2(0)\\
	&R\Psi^{\prime}_2(d/2)+\widetilde{\Psi}^{\prime}_1(\pi/2)=\widetilde{\Psi}^{\prime}_2(0),
	\end{split}\\[10pt]
	\begin{split}
	&\Psi_3(-d/2)=\widetilde{\Psi}_2(\pi/2)=\widetilde{\Psi}_3(0)\\
	&R\Psi^{\prime}_3(-d/2)+\widetilde{\Psi}^{\prime}_3(0)=\widetilde{\Psi}^{\prime}_2(\pi/2),
	\end{split}\\[10pt]
	\begin{split}
	&\Psi_4(-d/2)=\widetilde{\Psi}_3(\pi/2)=\widetilde{\Psi}_4(0)\\
	&R\Psi^{\prime}_4(-d/2)+\widetilde{\Psi}^{\prime}_4(0)=\widetilde{\Psi}^{\prime}_3(\pi/2),
	\end{split}\label{eq:QPC4}
	\end{align}
	Due to the lattice periodicity, the Bloch boundary condition read
	\begin{equation}\label{eq:Bloch2d}
	\begin{aligned}
	&C_3^{(\text{i})}=e^{i K_x T}C_1^{(\text{i})}\\
	&C_4^{(\text{i})}=e^{i K_y T}C_2^{(\text{i})}
	\end{aligned}
	\end{equation}
	where $\mathbf{K} = (K_x, K_y)$ is the polariton envelope wave vector originating
	from the periodicity of the array and $T = d + 2R$ is the
	period of the array. Eqs.~(\ref{eq:QPC1})-(\ref{eq:Bloch2d}) form a set of linear equations for the variables $\{C_\text{m}^{(\text{i})}\}\cup\{ \widetilde{C}_{\widetilde{\text{m}}}^{(\text{i})}\}$. The secular equation in obtained by equating the determinant to zero. 
	\section{Zigzag chain}
	\label{appendix:zigzag_sys_derivation}
	For the junctions from (1) to (4) in Fig.~\ref{fig:SSH}(a) they read as follows
	\begin{align}
	\begin{split}
	&\Psi_1(d/2)=\widetilde{\Psi}_2(\alpha)=\widetilde{\Psi}_1(0)\\
	&R\Psi^{\prime}_1(d/2)+\widetilde{\Psi}^{\prime}_2(\alpha)=\widetilde{\Psi}^{\prime}_1(0),
	\end{split}\\[10pt]
	\begin{split}
	&\Psi_2(-d/2)=\widetilde{\Psi}_1(2\pi-\alpha)=\widetilde{\Psi}_2(0)\\
	&R\Psi^{\prime}_2(-d/2)+\widetilde{\Psi}^{\prime}_2(0)=\widetilde{\Psi}^{\prime}_1(2\pi-\alpha),
	\end{split}\\[10pt]
	\begin{split}
	&\Psi_2(d/2)=\widetilde{\Psi}_4(2\pi-\alpha)=\widetilde{\Psi}_3(0)\\
	&R\Psi^{\prime}_2(d/2)+\widetilde{\Psi}^{\prime}_4(2\pi-\alpha)=\widetilde{\Psi}^{\prime}_3(0),
	\end{split}\\[10pt]
	\begin{split}
	&\Psi_3(-d/2)=\widetilde{\Psi}_3(\alpha)=\widetilde{\Psi}_4(0)\\
	&R\Psi^{\prime}_3(-d/2)+\widetilde{\Psi}^{\prime}_4(0)=\widetilde{\Psi}^{\prime}_3(\alpha).
	\end{split}
	\end{align}
	Due to the chain periodicity, the Bloch boundary condition read
	\begin{equation}
	C_3^{(i)}=e^{i K T}C_1^{(i)}.
	\end{equation}
	The secular equation is found by setting the determinant to zero. 
	
	\section{S-matrix approach}
	\label{appendix:s-matrix}
	In this section, we develop the theory of scattering matrices in the context of quantum rings. Buttiker et al.~\cite{Buttiker} considered the quantum ring connected to the current lead and assumed the scattering matrix $S$ to be symmetric with respect to the two arms of the ring and have shown that the $S$-matrix can be parameterized by only one real variable ${0\le\varepsilon^{\prime}\le1/\sqrt{2}}$ and has the following form
	\begin{equation}
	S(\varepsilon^{\prime})= 
	\begin{pmatrix}   
	-(a(\varepsilon^{\prime})+b(\varepsilon^{\prime})) &\varepsilon^{\prime}&\varepsilon^{\prime}\\
	\varepsilon^{\prime}&a(\varepsilon^{\prime})&b(\varepsilon^{\prime})\\
	\varepsilon^{\prime}&b(\varepsilon^{\prime})&a(\varepsilon^{\prime})\\
	\end{pmatrix}.
	\end{equation}
	Probability current conservation requires $S^{\dagger}=S$ or, in other words,
	\begin{align}\label{unitary_cond}
	&(a(\varepsilon^{\prime})+b(\varepsilon^{\prime}))^2+2\varepsilon^{\prime2}=1,\nonumber\\
	&a^2(\varepsilon^{\prime})+b^2(\varepsilon^{\prime})+\varepsilon^{\prime2}=1.
	\end{align}
	There are two types of solutions to Eqs.~(\ref{unitary_cond}) for the functions $a(\varepsilon^{\prime})$ and $b(\varepsilon^{\prime})$:
	\begin{align}
	a_{1,\pm}(\varepsilon^{\prime})=\pm\frac{1}{2}(\sqrt{1-2\varepsilon^{\prime2}}-1),\\
	b_{1,\pm}(\varepsilon^{\prime})=\pm\frac{1}{2}(\sqrt{1-2\varepsilon^{\prime2}}+1),
	\end{align}
	and
	\begin{align}
	a_{2,\pm}(\varepsilon^{\prime})=\pm\frac{1}{2}(\sqrt{1-2\varepsilon^{\prime2}}+1),\\
	b_{2,\pm}(\varepsilon^{\prime})=\pm\frac{1}{2}(\sqrt{1-2\varepsilon^{\prime2}}-1).
	\end{align}
	Thus, there exists four different types of the scattering matrices, namely, $S_{1,\pm}(\varepsilon^{\prime})$ and $S_{2,\pm}(\varepsilon^{\prime})$. 
	Here we show that the four types of the scattering matrices can be combined into two by the following procedure. We introduce  analytical single-valued functions $S_{\pm}(\varepsilon)$ of a complex argument $\varepsilon\in\mathbb{C}\backslash[-1/\sqrt{2},1/\sqrt{2}]$. They are defined as to satisfy
	\begin{equation}
	\lim\limits_{\varepsilon^{\prime\prime}\to+0} S_{\pm}(\varepsilon^{\prime}+i\varepsilon^{\prime\prime})=S_{1,\pm}(\varepsilon^{\prime})
	\end{equation}
	for $\varepsilon^{\prime}\in[0,1/\sqrt{2}]$, which, in turn, supplemented by the analyticity condition, specifies $S_{\pm}(\varepsilon)$ unambiguously. An explicit form of $S_{\pm}(\varepsilon)$ reads
	\begin{equation}
	S_{\pm}(\varepsilon)=
	\begin{pmatrix}   
	-(a_{\pm}(\varepsilon)+b_{\pm}(\varepsilon)) &\varepsilon&\varepsilon\\
	\varepsilon&a_{\pm}(\varepsilon)&b_{\pm}(\varepsilon)\\
	\varepsilon&b_{\pm}(\varepsilon)&a_{\pm}(\varepsilon)\\
	\end{pmatrix},
	\end{equation}
	where
	\begin{align}
	a_{\pm}(\varepsilon)&=\pm\frac{1}{2}(\sqrt{1-2\varepsilon^2}-1),\\
	b_{\pm}(\varepsilon)&=\pm\frac{1}{2}(\sqrt{1-2\varepsilon^2}+1)
	\end{align}
	are single-valued analytic functions of a complex argument $\varepsilon\in\mathbb{C}\backslash[-1/\sqrt{2},1/\sqrt{2}]$ 
	and we chose the branch of $\sqrt{1-2\varepsilon^2}$ to satisfy  
	\begin{equation}
	\lim\limits_{\varepsilon^{\prime\prime}\to+0}\sqrt{1-2(\varepsilon^{\prime}\pm i\varepsilon^{\prime\prime})^2}=\pm\sqrt{1-2\varepsilon^{\prime2}},
	\end{equation}
	where in the RHS the square root denotes a real-valued function of the real variable $\varepsilon^{\prime}\in[0,1/\sqrt{2}]$. From the definition of $S_{
		\pm}(\varepsilon)$ we have
	\begin{equation}
	S_{2,\pm}(\varepsilon^{\prime})=\lim\limits_{\varepsilon^{\prime\prime}\to+0}S_{\mp}(\varepsilon^{\prime}-i\varepsilon^{\prime\prime}).
	\end{equation}
	\begin{figure}[ht]
		\includegraphics[width=0.3 \textwidth]{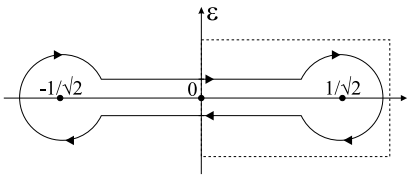}
		\caption{A branch cut in the complex plane for the analytic functions $S_{\pm}(\varepsilon)$ of $\varepsilon\in\mathbb{C}\backslash[-1/\sqrt{2},1/\sqrt{2}]$ and the contour $\mathcal{C}$ around the cut. The contour is divided into two parts $\mathcal{C}=\mathcal{C}_L\bigcup\mathcal{C}_R$, where $\mathcal{C}_R$ is marked by enclosing it in a rectangular box.} 
		\label{fig:s_matrix_contour}
	\end{figure}
	Thus, we have expressed $S_{1,\pm}(\varepsilon)$ and $S_{2,\pm}(\varepsilon)$ via the analytic functions $S_{\pm}(\varepsilon)|_{\varepsilon\in\mathcal{C}_R}$ on the contour $\mathcal{C}_R$, depicted in Fig.~\ref{fig:s_matrix_contour}. We focus on $S_{+}(\varepsilon)|_{\varepsilon\in\mathcal{C}_R}$ as it corresponds to the case when the transmitted waves have the same phase as the incident ones.
	
	Let us discuss the obtained result by tracing $S_{-}(\varepsilon)$ as we move along $\mathcal{C}_R$.   At the initial point of the contour $\mathcal{C}_R$ the $S$-matrix is 
	\begin{equation}
	\lim\limits_{\varepsilon^{\prime\prime}\to+0}S_{+}(0+i\varepsilon^{\prime\prime})=S_{1,+}(0)
	\end{equation}
	and a wave coming from the lead is totally reflected, while the waves in the ring do not see the junction. Following the directions, indicated by arrows in Fig.~\ref{fig:s_matrix_contour}, we arrive at the vicinity of $\varepsilon=1/\sqrt{2}$ point where the $S$-matrix is \begin{align}
	\lim\limits_{\varepsilon^{\prime\prime}\to0}S_{+}(1/\sqrt{2}+i\varepsilon^{\prime\prime})&=S_{1,+}(1/\sqrt{2})\\
	&=S_{2,-}(1/\sqrt{2})
	\end{align}
	We finish the route along the contour $\mathcal{C}_R$ on the lower bank of the branch cut, where we have
	\begin{equation}
	\lim\limits_{\varepsilon^{\prime\prime}\to+0}S_{+}(0-i\varepsilon^{\prime\prime})=S_{2,-}(0)
	\end{equation}
	which corresponds to an absolutely opaque lead-ring junction. 
	
	Alternatively, one could impose the continuity condition at the junction in addition to the current conservation (together they are usually referred to as Griffith's conditions). This would lead to some scattering matrix that we denote as $S_{\text{Griffith}}$. It turns out that 
	\begin{equation}
	S_{\text{Griffith}}=\lim\limits_{\varepsilon^{\prime\prime}\to+0}S_{+}(2/3+i\varepsilon^{\prime\prime})=S_{1,+}(2/3),
	\end{equation}
	which can be verified by a direct calculation.

    \bibliographystyle{apsrev4-1}
	\bibliography{PLbibliography}
\end {document}